\documentclass[twocolumn]{aastex63}

\usepackage{amsmath}
\usepackage{amssymb}
\usepackage{graphicx}

\begin{document}

\title{Influence of Star Cluster Mass, Age, and Galaxy Star Formation Rate on Star Cluster Radii}

\author{Michelle Sebastian}
\affiliation{Ventura County Community College District, Ventura County, CA, USA}


\begin{abstract}
Star clusters are key components of galaxies, and the relationship between cluster radius and mass encodes information about cluster formation and evolution. Theoretical models predict that age and specific star formation rate (sSFR) should influence cluster size through stellar mass loss and gas dynamics during formation. We hypothesized that if these theoretical predictions hold, multivariate models including age and sSFR should predict cluster radius better than models using mass alone. To test this, we used regression analysis on 5,105 star clusters from the LEGUS survey, comparing a full multivariate model against a mass-only baseline. We found that mass dominated the radius-mass relation: $\log(\mathrm{Mass})$ showed a strong correlation with radius (coefficient $= 0.131 \pm 0.008$, $p < 0.001$), while $\log(\mathrm{sSFR})$ and $\log(\mathrm{Age})$ contributed negligibly ($0.0002 \pm 0.015$ and $0.038 \pm 0.006$, respectively). Cross-validation revealed that the mass-only model generalized better (CV $R^2 = 0.028$ vs $-0.017$), with the negative value for the multivariate model indicating overfitting. Contrary to our hypothesis, adding age and sSFR did not improve predictive performance. The low $R^2$ (0.115) indicated that most variance in cluster radius remained unexplained by these variables, suggesting other factors may play important roles. Among the variables tested, our findings were consistent with virial equilibrium predictions, with mass serving as a more fundamental parameter than evolutionary age or galaxy star formation rate.
\end{abstract}

\section{Introduction} \label{sec:intro}

Star clusters are key components of galaxies and serve as natural laboratories for studying stellar evolution, dynamical processes, and galaxy formation history \citep{Krumholz2019}. The relationship between cluster radius and mass has long been seen as a critical scaling relation that tells us about cluster formation and evolution \citep{Larsen2004}.

Previous studies have found consistent radius-mass relations across different cluster populations. \citet{Brown2021} looked at 6,097 young star clusters across 31 LEGUS galaxies and found $R_{\mathrm{eff}} \propto M^{0.24}$ with intrinsic scatter of 0.25 dex. Their work showed that the scaling exponent varies with age, with younger clusters (less than 10 Myr) having shallower slopes.

The theoretical basis for the radius-mass relation comes from virial equilibrium---the balance between gravitational self-binding and internal kinetic energy \citep{Gieles2011}. For a system in virial equilibrium, $2K + U = 0$, where $K$ is kinetic energy and $U$ is gravitational potential energy. This basic relationship gives $M \propto \sigma^2 R$ for mass $M$, velocity dispersion $\sigma$, and radius $R$, predicting that mass and radius are closely linked. But theoretical models also predict other effects. Stellar evolution causes mass loss and possible radius expansion with age \citep{Alexander2014}. Active star formation and the gas dynamics that come with it might also affect how compact clusters are \citep{Krause2020}. High-pressure environments with rapid star formation could produce tighter clusters, or alternatively, turbulent support could lead to more spread-out clusters---theory predicts some effect, though the direction is uncertain \citep{Krause2020}.

Previous studies have set up univariate scaling relations and documented how cluster properties change with age \citep{Brown2021,Ryon2017}. But an important gap remained: few studies had actually tested whether age and sSFR improve radius predictions when combined with mass in multivariate models. Most work has looked at one variable at a time or reported qualitative trends without careful statistical testing of how multiple parameters work together. This meant we could not distinguish whether the simple mass-radius relation reflects real physical dominance of mass or just incomplete analysis.

Based on theoretical predictions from stellar evolution and star formation physics, we hypothesized that age and sSFR would improve predictions of cluster radius when combined with mass, resulting in superior predictive performance over mass-only models. We tested this hypothesis on 5,105 star clusters from the LEGUS survey using multivariate regression, variance inflation factor analysis, cross-validation, and information criteria. Contrary to our hypothesis, mass dominated the radius-mass relation, and the mass-only model showed better generalization despite the multivariate model fitting the training data slightly better.

\section{Materials and Methods} \label{sec:methods}

\subsection{Dataset} \label{sec:dataset}

Star cluster data from the LEGUS survey \citep{Calzetti2015} catalog compiled by \citet{Brown2021}, comprising 6,097 clusters from 31 nearby galaxies, were analyzed. The dataset included cluster masses ranging from $10^3$ to $10^7~M_\odot$, ages from 1 Myr to 12,000 Myr, and effective radii measured from HST imaging. Specific star formation rates were calculated as $\mathrm{SFR}/M_{\mathrm{stellar}}$, with SFR derived from UV and H$\alpha$ fluxes using standard calibrations \citep{Cook2019}. Only clusters with reliable radius measurements (quality flag = 1) and mass uncertainties $< 0.5$ dex were retained, yielding 5,105 clusters for analysis.

\subsection{Statistical Approach} \label{sec:stats}

Two regression models were built and compared. The full multivariate model used mass, age, and galaxy sSFR as predictors:
\begin{equation}
\log_{10}(r_{\mathrm{eff}}) = \beta_0 + \beta_1\log_{10}(M) + \beta_2\log_{10}(\mathrm{sSFR}) + \beta_3\log_{10}(\mathrm{Age}) + \epsilon
\end{equation}
A reduced model used only mass to set a baseline:
\begin{equation}
\log_{10}(r_{\mathrm{eff}}) = \beta_0 + \beta_1\log_{10}(M) + \epsilon
\end{equation}
Variables were log-transformed to linearize relationships and stabilize variance across the wide parameter ranges \citep{Isobe1990}. Residual analysis was used to check model assumptions through residual standard error calculation. Q-Q plots were used to check for normality, and Breusch-Pagan tests were used to check for heteroscedasticity.

Variance Inflation Factors (VIF) were calculated to assess multicollinearity, where $\mathrm{VIF} = 1/(1-R^2)$ for each predictor $j$ \citep{Isobe1990}. Values above 10 would indicate severe multicollinearity.

Multiple approaches were used to compare models. Training performance was measured using $R^2$, adjusted $R^2$, and root mean squared error (RMSE). Generalization was tested using 5-fold cross-validation, where data were split into 5 equal parts, with training on 4 parts and testing on the fifth, repeated 5 times with different test sets. AIC and BIC were calculated for model selection \citep{Burnham2002}. A partial $F$-test was conducted to assess whether adding sSFR and age significantly improved the mass-only model.

\subsection{Implementation} \label{sec:implementation}

All analyses were conducted in Google Colab using Python 3.9. Libraries used included pandas (v1.3.5), numpy (v1.21.5), scikit-learn (v1.0.2), statsmodels (v0.13.2), and matplotlib/seaborn for visualization. Code is available at \url{https://github.com/michelle-sebastian/starcluster}.

\section{Results} \label{sec:results}

\subsection{Mass Dominates the Radius-Mass Relation} \label{sec:mass_dominates}

To test whether age and sSFR improve radius predictions beyond mass alone, we first fitted a multivariate regression model including all three predictors, allowing us to quantify each variable's relative contribution, while controlling for the others. The full model including $\log(\mathrm{Mass})$, $\log(\mathrm{sSFR})$, and $\log(\mathrm{Age})$ as predictors explained only 12.2\% of radius variance ($R^2 = 0.122$), indicating that most variation in cluster radius remained unexplained by these variables. The fitted equation was:
\begin{equation}
\log_{10}(R_{\mathrm{eff}}) = -0.422 + 0.131 \times \log_{10}(M) + 0.0002 \times \log_{10}(\mathrm{sSFR}) + 0.038 \times \log_{10}(\mathrm{Age})
\end{equation}

Individual regression coefficients revealed major differences in predictor importance. Mass emerged as the dominant predictor, with a coefficient of 0.131 (SE $= 0.008$, $p < 0.001$), meaning that a tenfold increase in mass corresponds to a 35\% increase in radius (Table~\ref{tab:multivariate}). In contrast, sSFR showed an essentially zero coefficient of 0.0002 (SE $= 0.015$, $p = 0.99$), indicating no meaningful contribution to radius prediction. Age showed a statistically significant but small coefficient of 0.038 (SE $= 0.006$, $p < 0.001$), roughly 3.4 times weaker than mass (Table~\ref{tab:multivariate}).

\begin{deluxetable}{lccc}
\tablecaption{Regression coefficients for the full multivariate model \label{tab:multivariate}}
\tablehead{
\colhead{Parameter} & \colhead{Coefficient ($\beta$)} & \colhead{Std.\ Error} & \colhead{$p$-value}
}
\startdata
Intercept & $-0.422$ & 0.148 & 0.004 \\
$\log_{10}(\mathrm{Mass})$ & 0.131 & 0.008 & $<0.001$ \\
$\log_{10}(\mathrm{sSFR})$ & 0.0002 & 0.015 & 0.989 \\
$\log_{10}(\mathrm{Age})$ & 0.039 & 0.006 & $<0.001$ \\
\enddata
\tablecomments{The model used $\log_{10}(\mathrm{Mass})$, $\log_{10}(\mathrm{sSFR})$, and $\log_{10}(\mathrm{Age})$ as predictors of $\log_{10}(R_{\mathrm{eff}})$. Mass dominates with a coefficient of 0.131 ($p < 0.001$), while sSFR contributes essentially nothing (0.0002, $p = 0.989$), and age shows minimal though statistically significant contribution (0.039, $p < 0.001$). Sample size $n = 5{,}105$ clusters. Coefficients estimated using ordinary least squares regression in Python (statsmodels v0.13.2).}
\end{deluxetable}

\subsection{Multicollinearity Assessment} \label{sec:multicollinearity}

Before interpreting these coefficients, we checked whether correlations among predictors might inflate coefficient uncertainties or obscure true relationships. This step was necessary because astrophysical variables often correlate due to shared evolutionary pathways. Variance Inflation Factor (VIF) analysis showed all values fell well below the conventional threshold of 10, with the highest VIF of 1.56 indicating only minimal coefficient variance inflation (Table~\ref{tab:vif}). The correlation matrix confirmed moderate correlations: Mass-Age showed $r = 0.59$, Mass-sSFR showed $r = -0.09$, and Age-sSFR showed $r = 0.03$ (Table~\ref{tab:vif}). The moderate Mass-Age correlation likely reflects observational selection effects where massive clusters survive longer due to enhanced gravitational binding. These low VIF values confirmed that multicollinearity did not compromise our coefficient estimates.

\begin{deluxetable*}{lccccc}
\tablecaption{Multicollinearity diagnostics \label{tab:vif}}
\tablehead{
\colhead{Variable} & \colhead{VIF} & \colhead{Assessment} & \colhead{Corr.\ w/ $\log_{10}(\mathrm{Mass})$} & \colhead{Corr.\ w/ $\log_{10}(\mathrm{sSFR})$} & \colhead{Corr.\ w/ $\log_{10}(\mathrm{Age})$}
}
\startdata
$\log_{10}(\mathrm{Mass})$ & 1.56 & No concern & 1.000 & $-0.090$ & 0.590 \\
$\log_{10}(\mathrm{sSFR})$ & 1.02 & No concern & $-0.090$ & 1.000 & 0.034 \\
$\log_{10}(\mathrm{Age})$ & 1.55 & No concern & 0.590 & 0.034 & 1.000 \\
\enddata
\tablecomments{Variance Inflation Factor (VIF) values and correlation matrix for all predictors in the multivariate model. All VIF values fall well below the conventional threshold of 10, indicating no severe multicollinearity. The moderate Mass-Age correlation ($r = 0.59$) likely reflects observational selection effects where massive clusters survive longer due to enhanced gravitational binding. Sample size $n = 5{,}105$ clusters.}
\end{deluxetable*}

\subsection{Mass-Only Model as Baseline} \label{sec:massonly}

\begin{figure}[ht!]
\plotone{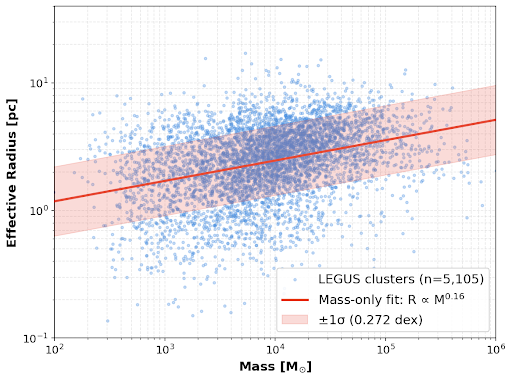}
\caption{Star cluster radius-mass relation using the mass-only model. Scatter plot showing effective radius ($R_{\mathrm{eff}}$) versus mass for 5,105 LEGUS star clusters analyzed using ordinary least squares regression in Python (statsmodels v0.13.2). The red line shows the best-fit relation $\log_{10}(R_{\mathrm{eff}}) = -0.247 + 0.160 \times \log_{10}(M)$, with shaded regions indicating $\pm 1\sigma$ (dark) and $\pm 2\sigma$ (light) confidence intervals. The model yields $R^2 = 0.115$ and residual standard error $= 0.272$ dex. Cluster data from \citet{Brown2021} were filtered to include only clusters with reliable radius measurements (quality flag = 1) and mass uncertainties $< 0.5$ dex.}
\label{fig:massonly}
\end{figure}

To test whether age and sSFR improve predictions beyond mass, we fitted a mass-only model as a baseline for comparison. The mass-only model yielded $\log_{10}(R_{\mathrm{eff}}) = -0.247 + 0.160 \times \log_{10}(M)$, indicating a shallow power-law scaling where $R_{\mathrm{eff}} \propto M^{0.16}$. Despite using only a single predictor, this model achieved $R^2 = 0.115$ with a residual standard error of 0.272 dex, capturing 94\% of the full model's explanatory power (Table~\ref{tab:massonly}, Figure~\ref{fig:massonly}).

\begin{deluxetable}{lccc}
\tablecaption{Regression coefficients for the mass-only model \label{tab:massonly}}
\tablehead{
\colhead{Parameter} & \colhead{Coefficient ($\beta$)} & \colhead{Std.\ Error} & \colhead{$p$-value}
}
\startdata
Intercept & $-0.247$ & 0.024 & $<0.001$ \\
$\log_{10}(\mathrm{Mass})$ & 0.160 & 0.005 & $<0.001$ \\
\enddata
\tablecomments{The model used only $\log_{10}(\mathrm{Mass})$ as a predictor of $\log_{10}(R_{\mathrm{eff}})$, yielding a power-law scaling where $R_{\mathrm{eff}} \propto M^{0.16}$. Sample size $n = 5{,}105$ clusters. Coefficients estimated using ordinary least squares regression in Python (statsmodels v0.13.2).}
\end{deluxetable}

\subsection{Residual Analysis Validates Model Assumptions} \label{sec:residuals}

\begin{figure}[ht!]
\plotone{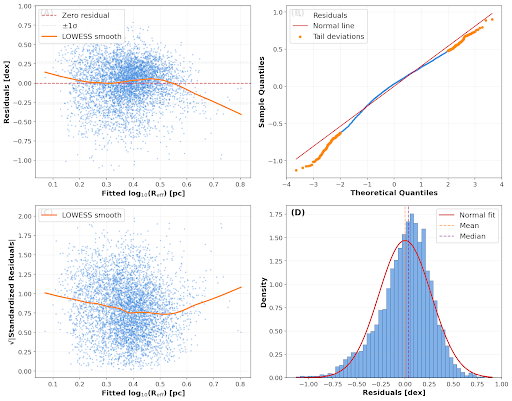}
\caption{Residual analysis for the mass-only model. Diagnostic plots generated using Python (matplotlib/seaborn) to assess regression assumptions for 5,105 LEGUS clusters. (A) Residuals versus fitted values showing no systematic bias; the orange LOWESS smooth line remains near zero across the fitted range, confirming model adequacy. The dashed lines indicate $\pm 1\sigma$ bounds. (B) Normal Q-Q plot showing approximate normality of residuals with correlation $= 0.992$ between observed and theoretical quantiles; heavy-tailed deviations are highlighted in orange. (C) Scale-location plot for heteroscedasticity assessment, with LOWESS smooth showing mild increase in variance at higher fitted values (Breusch-Pagan test $p < 0.001$). (D) Histogram of residuals showing symmetric distribution centered at zero (mean $= -0.0000$, median $= 0.0315$) with skewness $= -0.567$ and kurtosis $= 0.657$; the green curve shows a fitted normal distribution ($\sigma = 0.272$ dex).}
\label{fig:residuals}
\end{figure}

We examined residuals to verify that the mass-only model met regression assumptions, which is necessary for valid statistical inference. Residuals showed no systematic bias across the fitted range, scattering symmetrically around zero with a LOWESS smooth line remaining nearly horizontal (Figure~\ref{fig:residuals}A). The Q-Q plot indicated approximate normality with a correlation of 0.992 between observed and theoretical quantiles, though heavy tails suggested more outliers than a perfect normal distribution (Figure~\ref{fig:residuals}B). The Breusch-Pagan test indicated mild heteroscedasticity ($p < 0.001$), but the effect was subtle---residual variance increased only marginally with fitted values (Figure~\ref{fig:residuals}), a common characteristic of astronomical scaling relations spanning wide dynamic ranges \citep{Isobe1990}.

\subsection{Simpler Model Generalizes Better} \label{sec:generalization}

The critical test of our hypothesis compared generalization performance between models. If age and sSFR provide meaningful predictive improvement, the full model should predict new data better than the mass-only model. We used five-fold cross-validation, which trains models on 80\% of data and tests predictions on the held-out 20\%, repeating this five times to assess how well each model generalizes.

Cross-validation favored the simpler model. The mass-only model achieved CV $R^2 = 0.028$, while the full multivariate model achieved CV $R^2 = -0.017$ (Table~\ref{tab:comparison}). The negative cross-validation $R^2$ for the multivariate model pointed to overfitting---it performed worse than simply predicting the mean value for all clusters.

We also compared models using information criteria, which penalize complexity to guard against overfitting. Both AIC (Akaike Information Criterion) and BIC (Bayesian Information Criterion) favored the full model ($\Delta\mathrm{AIC} = -37.2$, $\Delta\mathrm{BIC} = -24.2$), representing ``very strong evidence'' by conventional guidelines (Table~\ref{tab:comparison}). However, this conflicted with cross-validation results.

A partial $F$-test comparing nested models yielded $F(2, 5101) = 20.69$, $p < 0.001$, indicating that adding sSFR and age provided statistically significant improvement over the mass-only model in training data. However, this 0.7\% improvement in $R^2$ did not translate to better generalization, as demonstrated by the negative cross-validation $R^2$ for the full multivariate model.

\begin{deluxetable*}{lcccc}
\tablecaption{Model performance comparison showing cross-validation favors mass-only model \label{tab:comparison}}
\tablehead{
\colhead{Metric} & \colhead{Multivariate Model} & \colhead{Mass-Only Model} & \colhead{Difference} & \colhead{Interpretation}
}
\startdata
\multicolumn{5}{c}{\textit{Training Performance}} \\
$R^2$ & 0.122 & 0.115 & 0.007 & Full model explains 6\% more variance \\
Adjusted $R^2$ & 0.121 & 0.115 & 0.006 & Minimal improvement after adjusting \\
RMSE & 0.271 & 0.272 & $-0.001$ & Essentially equivalent error \\
\hline
\multicolumn{5}{c}{\textit{Generalization Performance}} \\
CV $R^2$ (5-fold) & $-0.017$ & 0.028 & $-0.045$ & Mass-only generalizes much better \\
CV RMSE & 0.283 & 0.276 & 0.007 & Lower out-of-sample error for mass-only \\
\hline
\multicolumn{5}{c}{\textit{Information Criteria}} \\
AIC & $-13340.0$ & $-13302.8$ & $-37.2$ & Full model preferred by AIC \\
BIC & $-13313.9$ & $-13289.7$ & $-24.2$ & Full model preferred by BIC \\
\hline
\multicolumn{5}{c}{\textit{Statistical Tests}} \\
Partial $F$-test & $F(2,5101) = 20.69$ & --- & $p < 0.001$ & Statistically significant but small effect \\
\enddata
\tablecomments{Comparison of the full multivariate model (including mass, age, and sSFR) against the mass-only model across training performance, generalization performance, and information criteria metrics. $R^2$ represents the coefficient of determination measuring variance explained in training data. RMSE (Root Mean Square Error) quantifies average prediction error in dex. CV $R^2$ (Cross-Validation coefficient of determination) measures predictive performance on held-out data using 5-fold cross-validation, where negative values indicate the model performs worse than predicting the mean. AIC (Akaike Information Criterion) and BIC (Bayesian Information Criterion) are information-theoretic measures where lower values indicate better model fit; $\Delta$AIC and $\Delta$BIC represent differences between models. Cross-validation performed using scikit-learn (v1.0.2) with 5 random folds.}
\end{deluxetable*}

\section{Discussion} \label{sec:discussion}

Our results showed that mass dominates the radius-mass relation in star clusters, providing no evidence to support our hypothesis. While mass showed strong predictive power, sSFR contributed essentially nothing, and age showed only a weak effect despite having statistical significance. The null result remains scientifically meaningful: even though theory predicts that age and sSFR should affect cluster structure \citep{Alexander2014,Krause2020}, these variables added very little once mass was accounted for.

The near-zero contribution of sSFR tells us that galaxy-level gas conditions leading to different star formation rates do not produce matching differences in cluster radius \citep{Krause2020}. This goes against models suggesting that high-pressure, actively star-forming environments should make more compact clusters. Similarly, the small age effect goes against models predicting significant radius evolution through stellar mass loss and dynamical relaxation \citep{Alexander2014}. Stellar evolution theory says that mass loss from supernovae and stellar winds should cause gravitational unbinding and radius expansion over time, yet we found only weak age dependence. Future work could investigate why these theoretically predicted evolutionary effects do not show up in the LEGUS sample---whether because of compensating physical mechanisms, observational selection effects, or gaps in current theoretical models.

The better cross-validation performance of the mass-only model supports choosing the simpler model. The gap between training performance (both models similar, $R^2 \sim 0.12$) and cross-validation performance (mass-only better) suggests overfitting in the full model. The negative CV $R^2$ indicates that the full model learned noise patterns specific to the training data that did not carry over to new observations \citep{Ivezic2014}. Interestingly, AIC and BIC favored the complex model, which would be ``strong evidence'' by textbook standards \citep{Burnham2002}. This conflict arises because information criteria evaluate in-sample fit while cross-validation tests out-of-sample prediction. For predictive applications, cross-validation provides more reliable guidance when these metrics disagree \citep{Burnham2002}.

These results have implications for both observational and theoretical cluster studies. The dominance of mass means that cluster sizes can be understood mainly through a single parameter, which simplifies theoretical models. This is consistent with the idea that virial equilibrium---the balance between gravitational self-binding and internal kinetic energy---plays a central role in governing cluster structure across different environments. The observed mass coefficient of 0.131, meaning $R \propto M^{0.13}$, falls below the $R \propto M^{0.5}$ expected from pure virial scaling with constant velocity dispersion \citep{Gieles2011}. This difference could be due to observational selection effects, tidal truncation, and the mix of evolutionary states in our sample.

This study has some limitations. The LEGUS sample, while large, covers nearby galaxies that may not capture the full range of cluster environments. The moderate mass-age correlation ($r = 0.59$) likely reflects observational selection where old, low-mass clusters fall below detection limits, possibly hiding age effects present in the full cluster population. Future work could look at environmental parameters like gas density and tidal field strength that may affect the mass-radius relation, as well as consider employing machine learning methods to find non-linear patterns that standard regression might miss.

In summary, we tested whether theoretical predictions of age and galaxy sSFR effects would translate to better predictive power for cluster radius. We found that mass is the main parameter controlling star cluster radius, with age and sSFR adding no meaningful predictive improvement despite theoretical expectations. The mass-only model captured 94\% of the full model's explanatory power while generalizing better, and the negative cross-validation $R^2$ for the multivariate model showed potential overfitting. These findings are consistent with virial equilibrium physics playing a central role in setting cluster structure.

\acknowledgments

We thank Gillen Brown, PhD, for guidance and constructive feedback on this research and manuscript. This research made use of publicly available star cluster catalogs from the LEGUS survey and curated data from \citet{Brown2021}.

\end{document}